\documentclass[12pt]{article}

\usepackage{amsmath,amsthm,amssymb,amsopn,amsfonts}
\usepackage{graphics}
\usepackage{graphicx}
\usepackage{fullpage}

\newcommand{\twospace}{\renewcommand{\baselinestretch}{1.3}\normalsize}

\newcommand{\Pry}{\mathbb{P}}
\newcommand{\X}{{\mathcal X}}
\newcommand{\Y}{{\mathcal Y}}
\newcommand{\UU}{{\mathcal U}}
\newcommand{\VV}{{\mathcal V}}
\newcommand{\R}{\mathbb{R}}

\newtheorem{theorem}{Theorem}

\newtheorem{note}[theorem]{Note}

\newtheorem{lemma}[theorem]{Lemma}
\newtheorem{corollary}[theorem]{Corollary}

\twospace

\begin{document}
\title{Consistent adjacency-spectral partitioning for the
stochastic block model when the model parameters are unknown}
\author{Donniell E. Fishkind, Daniel L. Sussman, Minh Tang, Joshua T. Vogelstein\\and Carey E. Priebe\\
Department of Applied Mathematics and Statistics, Johns Hopkins University}

\maketitle
\begin{abstract}
For random graphs distributed according to a stochastic block model, we consider the inferential task of partioning vertices into blocks using spectral techniques.
Spectral partioning using the normalized Laplacian and the adjacency matrix have both been shown to be consistent as the number of vertices tend to infinity. % in the setting that the random graph is distributed according to a stochastic block mode.  
Importantly, both procedures require that the number of blocks and the rank of the communication probability matrix are known, even as the rest of the parameters may be unknown.
In this article, we prove that the (suitably modified) adjacency-spectral partitioning procedure, requiring only an upper bound on the rank of the communication probability matrix, is consistent.
Indeed, this result demonstrates a robustness to model mis-specification; an overestimate of the rank may impose a moderate performance penalty, but the procedure is still consistent.
Furthermore, we extend this procedure to the setting where adjacencies may have multiple modalities and we allow for either directed or undirected graphs.

%A stochastic block model consists of a random partition
%of $n$ vertices into blocks $1,2,\ldots,K$ for which, conditioned on the partition,
%every pair of vertices has probability of adjacency entirely
%determined by the block membership of the two vertices. The model parameters are~$K$, the distribution
%of the random partition, and a {\it communication probability matrix} $M$ in
%$[0,1]^{K \times K}$ listing the adjacency probabilities associated with all pairs of blocks.
%Suppose a realization of the $n \times n$ vertex adjacency matrix
%is observed, but the underlying
%partition of the vertices into blocks is not observed; the main inferential
%task is to correctly partition the vertices into the blocks with
%only a negligible number of vertices misassigned.
%
%
%%For this inferential task, 
%Rohe et al.~\cite{RCY} prove the consistency of spectral partitioning
%applied to the normalized Laplacian, and Sussman et al.~\cite{STFP} extend this to prove consistency of spectral partitioning directly on the adjacency matrix. 
%Importantly, both procedures require that $K$ and rank$M$ are known, even as the rest of the parameters may be unknown. 
%In this article, we prove that the (suitably modified) adjacency-spectral partitioning procedure, requiring only an upper bound on rank$M$, is consistent.
%Indeed, this result demonstrates a robustness to model mis-specification; an overestimate of rank$M$ imposes a moderate penalty in the bounds but the procedure is still consistent. 
%%In particular, this result shows a robustness in the adjacency-spectral partitioning procedure.
\end{abstract}

\newpage

\section{Background and overview}

Our setting is the {\it stochastic block model} \cite{HLL,WW}---a random graph
model in which a set of $n$ vertices is randomly partitioned into $K$
{\it blocks} and then, conditioned on the partition, existence of edges
between all pairs of vertices are independent Bernoulli trials with
parameters determined by the block membership of the pair.
(The model details are specified in Section~\ref{c}.)

The realized partition of the vertices is not observed, nor are the Bernoulli trial
parameters known. However, the realized vertex
adjacencies (edges) are observed, and the main
inferential task is to estimate the partition of the vertices, using the realized
adjacencies as a guide. Such an estimate will be called consistent if and
when, in considering a sequence of realizations for $n=1,2,3,\ldots$ with
common model parameters, it happens almost surely that the fraction of misassigned vertices converges to zero as $n \rightarrow \infty$.

Rohe et al.~\cite{RCY} proved the consistency of
a block estimator that is based on spectral partitioning
applied to the normalized Laplacian, and Sussman et al.~\cite{STFP} extended this to prove the consistency of a block estimator that is based on
spectral partitioning applied to the adjacency matrix. 
Importantly, both of these
procedures assume that $K$ and the rank of $M$  are known
(where $M \in [0,1]^{K \times K}$
is the matrix consisting of the Bernoulli parameters
for all pairs of blocks), even as the rest of
the parameters may be unknown. In this article, we prove that the
(suitably modified) adjacency-spectral partitioning procedure, requiring only an upper bound for rank$M$, gives consistent
block estimation. %even if the only thing that is known about the parameters is any upper bound on the rank of $M$. (
We demonstrate a robustness to mis-specification of rank$M$; in particular, if a practitioner
overestimates the rank of $M$ in carrying out adjacency spectral
partitioning to estimate the blocks,
then the consistency of the procedure is not lost.
Indeed, this is a model selection result, and we provide estimators for $K$ and prove their consistency. %)

Our analysis and results are valid for both directed and undirected graphs.
We also allow for more than one modality of adjacency. For instance,
the stochastic block model can model a social network in which
the vertices are people, and the blocks are different
communities within the network such that probabilities of
communication between individual people are community dependent,
and there is available information about several different modes of
communication between the people; e.g.~who phoned whom on cell phones,
who phoned whom on land lines, who sent email to whom, who sent snail mail
to whom, with a separate adjacency matrix for each modality of communication.
Indeed, if there are different matrices $M$ for each mode of
communication, even if there is dependence in the communications
between two people across different modalities, our analysis and
results will hold---provided
that every pair of blocks is ``probabilistically discernable"
within at least one mode of communication.
(This will be made more precise in Section~\ref{c}.)

Latent space models (e.g.~Hoff et al.~\cite{HRH}) and, specifically,
random dot product models (e.g.~Young and Scheinerman \cite{YS}) give
rise to the stochastic block model. Indeed,
the techniques that we use in this article involve generating
latent vectors for a random dot product model structure which we then use
in our analysis. Nonetheless, our results can be used without awareness
of such random-dot-product-graph underlying structure, and we do not
concern ourselves here with estimating latent vectors for the blocks.
(In any event, latent vectors are not uniquely determinable here).

Consistent block estimation in stochastic block models has received much attention. Fortunato \cite{Fortunato} and Fjallstrom \cite{Fjallstrom} provide reviews of partitioning techniques for graphs in general.
Consistent partitioning of stochastic block models for two blocks
was accomplished by Snijders and Nowicki~\cite{SN} in 1997 and 
for equal-sized blocks by Condon and Karp \cite{CK} in 2001.
For the more general case, Bickel and Chen \cite{BC} in 2009 demonstrated a stronger version of consistency via
maximizing Newman-Girvan modularity \cite{NG} and other modularities.
For a growing number of blocks, Choi et al.~\cite{CWA} in 2010 proved consistency of likelihood based methods. In 2012, Bickel et al. \cite{BCL} provided a method to consistently estimate the stochastic block model parameters using subgraph counts and degree distributions. This work and the work of Bickel and Chen \cite{BC} both consider the case of very sparse graphs.

Rohe et al.~\cite{RCY} in 2011 used spectral partitioning on the
normalized Laplacian to consistently estimate a growing number of blocks and they allow the minimum expected degree to be at least $\Theta(n/\sqrt{\log n})$. Sussman
et al.~\cite{STFP} extended this
to prove consistency of spectral partitioning directly on the
adjacency matrix for directed and undirected graphs. Finally, Rohe et al.~\cite{Rohe2} proved consistency of bi-clustering on a directed version of the Laplacian for directed graphs. Unlike modularity and likelihood based methods, these spectral partitioning methods are computationally fast and easy to implement. Our work extends these spectral partitioning results to the situation when the number of blocks and the rank of the communication matrix is unknown. We present the situation for fixed parameters, and in Section~\ref{sec:disc} we discuss possible extensions.

The adjacency matrix has been previously used for
block estimation in stochastic block models by McSherry \cite{MS}, who proposed a randomized algorithm when the number of blocks as well as the block sizes are known.  Coja-Oghlan \cite{Coja} further investigate the methods proposed in McSherry and extend the work to sparser graphs. This method relies on bounds in the operator norm which have also been investigated by Oliveira \cite{Oliveira} and Chung et al. \cite{Chung}. In 2012, Chaudhuri et al. \cite{Chaudhuri} used an algorithm similar to the one in McSherry \cite{MS} to prove consistency  for the degree corrected planted partition model, a slight restriction of the degree corrected stochastic block model proposed in \cite{Karrer}. Notably, Chaudhuri et al. \cite{Chaudhuri} do not assume the number of blocks is known and provide an alternative method to estimate the number of blocks. This represents another important line of work for model selection in the stochastic block model.

The organization of the remainder of this article is as follows.
In Section~\ref{jjj} we describe the stochastic block model,
then we describe the inferential task and the adjacency-spectral
partitioning procedure for the task---when very little is known
about the parameters of the stochastic block model. In Section~\ref{kkk}
ancillary results and bounds are proven, followed  in Section~\ref{f}
by a proof of the consistency of our adjacency-spectral partitioning.
However, through Section~\ref{f}, there is an extra assumption that
the number of blocks $K$
is known. In Section~\ref{e} we provide a consistent
estimator for $K$, and in Section \ref{hhh} we prove the consistency of an
extended adjacency-spectral procedure that does not assume that $K$ is known.
Indeed, at that point, the only aspect of the model parameters
which is still assumed to be known is just an upper bound for the rank of
the communication probability matrix~$M$.

Bickel et al.~\cite{BCL} mention
the work of Rohe et al.~\cite{RCY}  as an important step, and
then opine that ``unfortunately
this does not deal with the problem [of] how to pick a
block model which is a good
approximation to the nonparametric model."
Taking these words to heart, our focus in this article is
on showing a robustness in the consistency of spectral
partitioning in the stochastic block model when using the adjacency matrix.
Our focus is on removing the need to know a priori the parameters,
and to still attain consistency in partitioning.
This robustness opens the door to explore
principled use of spectral techniques even for settings
where the stochastic block model assumptions do not
strictly hold, and we anticipate more future progress in consistency
results for spectral partitioning in nonparametric models.

We conclude the article with additional discussion of
consistent estimation of $K$ (Section~\ref{nnn}), illustrative
simulations (Section \ref{ppp}), and a brief discussion (Section~\ref{sec:disc}).

\section{The model, the adjacency-spectral partitioning procedure,
and its consistency  \label{jjj}  }

\subsection{The stochastic block model \label{c}}%%%%%%%%%%%%%%%%%

The random graph setting in which we work is the {\it stochastic block model},
which has parameters $K,\rho,M$ where
positive integer $K$  is the number of blocks,
 the {\it block probability vector} $\rho \in (0,1]^K$
satisfies $\sum_{k=1}^K \rho_k = 1$, and the {\it
communication probability matrix} $M \in [0,1]^{K \times K}$ satisfies
the model identifiability requirement that,
for all $p,q \in \{ 1,2,\ldots,K \}$ distinct, either it holds that
$M_{p,\cdot} \ne M_{q,\cdot}$ (i.e.~the $p$th and $q$th rows of $M$ are not equal)
or  $M_{\cdot,p} \ne M_{\cdot,q}$ (i.e.~the $p$th and $q$th columns of $M$ are not equal). The model is defined (and the parameters have roles) as follows:

There are $n$ vertices, labeled $1,2,\ldots,n$, and they are each
randomly assigned to blocks labeled $1,2,\ldots,K$
by a random {\it block membership function} $\tau :\{ 1,2,\ldots,n\}
\rightarrow \{ 1,2,\ldots,K \}$ such that for each vertex $i$ and block $k$,
independently of the other vertices, the probability that $\tau(i)=k$ is $\rho_k$.

Then there is a random adjacency matrix $A \in \{0,1\}^{n \times n}$
where, for all pairs of vertices $i,j$ that are distinct,
$A_{i,j}$ is $1$ or $0$ according
as there is an $i,j$ edge or not. Conditioned on $\tau$, the
probability of there being an $i,j$ edge is $M_{\tau(i),\tau(j)}$,
independently of the other pairs of vertices. Our analysis and results will
cover both the {\it undirected setting} in which edges are unordered pairs
(in particular, $A$ and $M$ are symmetric) and also the {\it directed setting}
in which edges are ordered pairs (in particular, $A$ and $M$ are not necessarily symmetric). In both settings the diagonals of $A$ are all $0$'s (i.e.~there are no
``loops" in the graph).

We assume that the parameters of the stochastic block model are not known,
except for one underlying assumption; namely,
that a positive integer $R$ is known that satisfies rank$M \leq R$.
(Of course, $R$ may be taken to be rank$M$ or $K$ if either of these
happen to be known.)
However, for now through Section \ref{f}, we also assume that $K$ is known;
in Section~\ref{e} we will provide a consistent estimator for $K$ if
$K$ is not known, and then in Section~\ref{hhh} we utilize this
consistent estimator for $K$ to extend all of the previous procedures
and results to the scenario where $K$ is also not known (and then the
only remaining assumption is our one underlying assumption
that a positive integer $R$ is known such
that rank$M \leq R$).

Although the realized adjacency matrix $A$ is observed,
the block membership function $\tau$ is not observed and,
indeed, the inferential task here is to estimate $\tau$.
In Section \ref{b}, adjacency-spectral partitioning is used to
obtain a {\it block {\bf assignment} function} $\hat{\tau}:\{1,2,\ldots,n\} \rightarrow
\{1,2,\ldots,K\}$ that serves as an estimator for $\tau$, up to permutation of the
block labels $1,2,\ldots,K$ on the $K$ blocks. Then Theorem~\ref{a} in Section \ref{d}
asserts that almost always the number of misassignments
$\min_{\textup{bijections }\pi: \{ 1,2,\ldots,K \}
\rightarrow \{ 1,2,\ldots,K \} } | \{ j=1,2,\ldots,n: \tau(j) \ne
\pi(\hat{\tau}(j))  \} |$ is negligible.

A more complicated scenario is where there are multiple ``modalities
of communication" for the vertices. Specifically, instead of one
probability communication matrix, there are
several probability communication matrices
$M^{(1)},M^{(2)}, \ldots, M^{(S)} \in [0,1]^{K \times K}$
which are all parameters of the model,
and there are corresponding random adjacency matrices
$A^{(1)},A^{(2)},\ldots,A^{(S)} \in \{ 0,1 \}^{n \times n}$ such that
for each {\it modality} $s=1,2,\ldots,S$ and for each pair of
vertices $i,j$ that are distinct, $A_{i,j}$ is $1$ with probability
$M^{(s)}_{\tau(i),\tau(j)}$ independently of the other pairs of vertices
but possibly with dependence across the modalities. As above,
for model identifiability purposes we assume that, for each $p,q \in \{1,2,\ldots,K \}$
distinct, there exists an $s \in \{1,2,\ldots,S\}$ such that $M^{(s)}_{p,\cdot}
\ne M^{(s)}_{q,\cdot}$ or $M^{(s)}_{\cdot,p} \ne M^{(s)}_{\cdot,q}$.
Also, it is assumed that we know positive integers
$R^{(1)},R^{(2)},\ldots,R^{(S)}$ which are upper bounds on 
$\mathrm{rank}M^{(1)},\mathrm{rank}M^{(2)}, \dotsc,\mathrm{rank}M^{(S)}$
%rank$M^{(1)}$, rank$M^{(2)}$, \ldots, rank$M^{(S)}$, 
respectively.
We will also describe next in Section~\ref{b} how the adjacency-spectral
partitioning procedure of that section can be modified
for this more complicated scenario so that Theorem \ref{a} will still hold for it.

\subsection{The adjacency-spectral partitioning procedure \label{b}}%%%%%%%%%%%%%

The adjacency-spectral partitioning procedure that we work with is given as follows:

First, take the realized adjacency matrix $A$, and
compute a singular value decomposition
 $A=[U | U_r ] (\Sigma \oplus \Sigma_r) [V|V_r]^T$ where
$U,V \in \R^{n \times R}$, $U_r,V_r \in \R^{n \times (n-R)}$,
$\Sigma\in\R^{R \times R}$, and $\Sigma_r~\in~\R^{(n-R) \times (n-R)}$ are
such that
$[U|U_r]$ and $[V|V_r]$ are each real-orthogonal matrices, and $\Sigma \oplus
\Sigma_r$ is a diagonal matrix with its diagonals non-increasingly ordered
$\sigma_1 \geq \sigma_2 \geq \sigma_3  \ldots  \geq \sigma_n$. Let
$\sqrt{\Sigma} \in \R^{R \times R}$ denote the diagonal matrix whose
diagonals are the nonnegative square roots of the respective diagonals~of~$\Sigma$,
and then compute $X:=U \sqrt{\Sigma}$ and $Y:=V \sqrt{\Sigma}$.

Then, cluster the rows of $X$ or $Y$ or $[X|Y]$ into at most $K$ clusters
 using the minimum least squares criterion, as follows:
If it is known that the rows of $M$ are pairwise not equal, then compute
${\mathcal C} \in \R^{n \times R}$ which minimizes $\| C-X \|_F$ over
all matrices $C \in \R^{n \times R}$ such that there are at most $K$ distinct-valued
rows in~$C$, otherwise, if it is known that the columns of $M$ are pairwise not
equal, then compute ${\mathcal C} \in \R^{n \times R}$ which
minimizes $\| C-Y \|_F$ over all matrices $C \in \R^{n \times R}$ such that
there are at most $K$ distinct-valued rows in $C$, otherwise compute
${\mathcal C} \in \R^{n \times 2R}$ which minimizes $\| C-[X|Y] \|_F$ over
all matrices $C \in \R^{n \times 2R}$ such that there are at most $K$ distinct-valued
rows in $C$. (Although our analysis will assume the use of this
minimum least squares criterion, note that popular clustering algorithms
such as $K$-means will also (empirically) produce good results for our
inferential task of block assignment.)

The clusters obtained are estimates for the
true blocks; i.e.~define the block assignment function $\hat{\tau}:
\{ 1,2,\ldots,n \} \rightarrow \{ 1,2,\ldots,K \} $ such that
the inverse images $\{ \hat{\tau}^{-1}(i):i=1,2,\ldots K \}$
partition the rows of ${\mathcal C}$ (by index)
so that rows in each part are equal-valued.
This concludes~the~procedure.

In the more complicated scenario of multiple modalities of communication,
carry out the above procedure in the same way, mutatis mutandis: For each modality
$s$, compute the singular value decomposition
$A^{(s)}=[U^{(s)} | U^{(s)}_r ] (\Sigma^{(s)} \oplus \Sigma_r^{(s)}) [V^{(s)}|V^{(s)}_r]^T$ for
$U^{(s)},V^{(s)} \in \R^{n \times R^{(s)}}$, $U_r^{(s)},V_r^{(s)}
 \in \R^{n \times (n-R^{(s)})}$,
$\Sigma \in \R^{R^{(s)} \times R^{(s)}}$,
and $\Sigma_r \in \R^{(n-R^{(s)}) \times (n-R^{(s)})}$ such that
$[U^{(s)}|U_r^{(s)}]$ and $[V^{(s)}|V_r^{(s)}]$ are each real-orthogonal matrices and $\Sigma^{(s)} \oplus
\Sigma_r^{(s)}$ is a diagonal matrix with its diagonals non-increasingly ordered,
then define $X^{(s)}:=U^{(s)} \sqrt{\Sigma^{(s)}}$ and
$Y^{(s)}:=V^{(s)} \sqrt{\Sigma^{(s)}}$ and then, according as
the rows of all $M^{(s)}$ are known to be distinct-valued, the columns of
$M^{(s)}$ are known to be distinct-valued, or neither,
compute ${\mathcal C}$ which minimizes $\| C -[X^{(1)}|X^{(2)}|
\cdots |X^{(S)}] \|_F$ or  $\| C -[Y^{(1)}|Y^{(2)}|
\cdots |Y^{(S)}] \|_F$  or $\| C -[X^{(1)}|X^{(2)}|
\cdots |X^{(S)}|Y^{(1)}|Y^{(2)}| \cdots |Y^{(S)} ] \|_F$ such that
there are at most $K$ distinct-valued rows in $C$, and then define
$\hat{\tau}$ as the partition of the vertices into
 $K$ blocks according to equal-valued corresponding rows in ${\mathcal C}$.

\subsection{Consistency of the adjacency-spectral partitioning of Section \ref{b} \label{d}}%%%%%%%%%%%%%

We consider a sequence of realizations of the stochastic block model given in
Section \ref{c} for successive values $n=1,2,3,\ldots$ with all stochastic block model parameters being fixed. In this article, an event will be said to hold
{\it almost always} if almost surely the event occurs for all
but a finite number of $n$.
The following consistency result asserts that
the number of misassignments in the adjacency-spectral procedure
of Section~\ref{b} is negligible; it will be proven in Section \ref{f}.

\begin{theorem} \label{a}
With the adjacency-spectral partitioning procedure of Section \ref{b},
for any fixed $\epsilon>\frac{3}{4}$,
the number of misassignments $\min_{\textup{bijections }\pi: \{ 1,2,\ldots,K \}
\rightarrow \{ 1,2,\ldots,K \} } | \{ j=1,2,\ldots,n: \tau(j) \ne
\pi(\hat{\tau}(j))  \} |$ is almost always less than $n^\epsilon$.
\end{theorem}

Theorem \ref{a} holds for all of the scenarios we described in Section \ref{c};
whether the edges are directed or undirected, whether there is one
modality of communication or multiple modalities. It also doesn't matter
if for each successive $n$ the partition function and adjacencies  are
re-realized for all vertices or if instead they are carried over
from previous $n$'s realization with just one new vertex randomly assigned
to a block and just this vertex's adjacencies to the previous vertices
being newly realized. (Note that if
the partition function and adjacencies
are re-realized for all vertices for successive $n$
then when we invoke the Strong
Law of Large Numbers we will be using the version of the Law
in \cite{HMT}.)

In Sussman et al.~\cite{STFP},
it was shown that if $R=$ rank$M$ then the number of misassignments
of the adjacency spectral procedure in Section \ref{b} is almost always less than a
constant times $\log n$ (where the constant is a function of the
model parameters). Indeed, both $\log n$ and $n^\epsilon$, when divided
by the number of vertices $n$, converge to zero, and in that sense we
can now say that whether rank$M$ is known or if it is overestimated
then either way the number of misassignments of spectral-adjacency
partitioning is negligible. This is a useful robustness result.

\section{Ancillary results  \label{kkk}}

\subsection{Latent vectors and constants from the model parameters \label{g}
}%%%%%%%%

In this section we identify relevant constants
$\alpha$, $\beta$, and $\gamma$ which depend
on the specific values of the
stochastic block model parameters; these constants will be used in our analysis.
 We also consider a particular decomposition of a model parameter
(the communication probability matrix~$M$) into {\it latent vectors}
which we may then usefully associate with the respective blocks.

We first emphasize that knowing the values of
these constants $\alpha$, $\beta$, and $\gamma$
which we are about to identify and knowing the values of the
latent vectors which we are about to define are
not at all needed to actually {\bf perform} the
adjacency-spectral clustering procedure
of Section \ref{b}, nor is any such knowledge needed in order
to invoke and {\bf use} the consistency result Theorem \ref{a}.
These constants and latent vectors will be used here
in developing the analysis and then proving Theorem~\ref{a}.

The stochastic block model parameters are $K$,\ $\rho$,\ $M$; the
constants $\alpha$, $\beta$, $\gamma$ are defined as follows:
Recall that
$\rho_k>0$ for all $k$; choose constant $\alpha >0$ such that
$\alpha < \rho_k$ for all $k$.
Next, choose matrices $\mu,\nu \in \R^{K
\times \textup{Rank}M}$ such that $M = \mu \nu^T$;
indeed, such matrices $\mu$ and $\nu$ (exist and) can be easily computed using a
singular value decomposition of $M$.
It is trivial to see that if any two rows
of $M$ are not equal-valued then those two corresponding
rows of $\mu$ must be not equal-valued,
and if any two columns of $M$ are not equal-valued then those two corresponding
rows of $\nu$ are not equal-valued. Choose constant $\beta >0$ be such that,
for all pairs of nonequal-valued rows  $\mu_{k,\cdot}$, $\mu_{k',\cdot}$ of $\mu$
it holds that $\|\mu_{k,\cdot}-\mu_{k',\cdot}\|_2>\beta$, and for all
pairs of nonequal-valued rows $\nu_{k,\cdot}$, $\nu_{k',\cdot}$ of $\nu$
it holds that $\|\nu_{k,\cdot}-\nu_{k',\cdot}\|_2>\beta$.
Lastly, since $\mu$ and $\nu$ are
full column rank, choose constant $\gamma>0$ such that the eigenvalues
of $\mu^T\mu$ and $\nu^T\nu$ are all greater than $\gamma$.

The rows of $\mu$ and $\nu$ are respectively called {\it left
latent vectors} and {\it right latent vectors}, and are associated
with the vertices as follows. The matrices ${\mathcal X} \in \R^{n \times
\textup{rank}M}$ and  ${\mathcal Y} \in \R^{n \times
\textup{rank}M}$ are defined such that for all $i=1,2,\ldots,n$,
${\mathcal X}_{i,\cdot}:=\mu_{\tau(i),\cdot}$ and ${\mathcal Y}_{i,\cdot}:=\nu_{\tau(i),\cdot}$. The significance of the latent
vectors is that for any pair of distinct vertices $i$ and $j$ the probability
of an $i,j$ edge is the inner product of the left latent vector
associated with $i$ (which is ${\mathcal X}_{i,\cdot}$) with the right latent vector
associated with $j$ (which is ${\mathcal Y}_{j,\cdot}$). Of course,
these latent vectors are not observed; indeed, $M$ is not known and
$\tau$ is not observed.

Finally, let
${\mathcal X}{\mathcal Y}^T={\mathcal U} \Lambda {\mathcal V}^T$
be a singular value decomposition, i.e.~${\mathcal U},{\mathcal V} \in \R^{n \times \textup{rank}M}$
each have orthonormal columns and $\Lambda \in \R^{\textup{rank}M
\times \textup{rank}M}$ is a diagonal matrix with diagonals ordered
in nonincreasing order $\varsigma_1 \geq \varsigma_2 \geq \varsigma_3 \geq
\cdots \geq \varsigma_{\textup{rank}M}$. It is useful to observe
that ${\mathcal X}({\mathcal Y}^T{\mathcal V}\Lambda^{-1})={\mathcal U}$
and $(\Lambda^{-1}{\mathcal U}^T{\mathcal X}){\mathcal Y}^T={\mathcal V}^T$
imply that rows of ${\mathcal X}$ which are equal-valued
correspond to rows of ${\mathcal U}$ that are equal-valued,
and rows of ${\mathcal Y}$ which are equal-valued correspond to rows of
${\mathcal V}$ that are equal-valued.

In the more complicated scenario of more than one communication modality
these definitions are made in the same way, mutatis mutandis:
For all modalities $s$, choose $\mu^{(s)},\nu^{(s)} \in \R^{K
\times \textup{Rank}M^{(s)}}$ such that
$M^{(s)}=\mu^{(s)}\nu^{(s)^T}$, then choose $\beta >0$
such that for every modality $s$ and all pairs of nonequal-valued
rows  $\mu^{(s)}_{k,\cdot}$, $\mu^{(s)}_{k',\cdot}$ of $\mu^{(s)}$
it holds that $\|\mu^{(s)}_{k,\cdot}-\mu^{(s)}_{k',\cdot}\|_2>\beta$, and for all
pairs of nonequal-valued
 rows  $\nu^{(s)}_{k,\cdot}$, $\nu^{(s)}_{k',\cdot}$ of $\nu^{(s)}$
it holds that $\|\nu^{(s)}_{k,\cdot}-\nu^{(s)}_{k',\cdot}\|_2>\beta$.
Choose constant $\gamma>0$ such that all  eigenvalues
of $\mu^{(s)^T}\mu^{(s)}$ and $\nu^{(s)^T}\nu^{(s)}$ for all modalities $s$
are greater than $\gamma$. Then, for each modality
$s$, define the rows of ${\mathcal X}^{(s)} \in \R^{n
\times \textup{Rank}M^{(s)}}$ and ${\mathcal Y}^{(s)} \in \R^{n
\times \textup{Rank}M^{(s)}}$ to be the rows from $\mu^{(s)}$ and $\nu^{(s)}$, respectively, corresponding to the blocks of the respective vertices,
and then define ${\mathcal U}^{(s)}$,
${\mathcal V}^{(s)}$, and $\Lambda^{(s)}$ (with
ordered diagonals $\varsigma_1^{(s)}, \varsigma_2^{(s)}, \ldots \varsigma^{(s)}_{\textup{rank}M^{(s)}}$) to form singular value decompositions
$\X^{(s)} \Y^{(s)^T}={\mathcal U}^{(s)} \Lambda^{(s)} {\mathcal V}^{(s)^T}$.

\subsection{Bounds}%%%%%%%%%%%%%%%%%%%%

In this section we prove a number of bounds involving
$A$, $\X \Y^T$, their singular values and
matrices constructed from components
of their singular value decompositions. These bounds
will then be used  in Section \ref{f} to prove Theorem \ref{a},
which asserts the consistency of the adjacency-spectral partitioning procedure of
Section \ref{b}.

The results in this section are stated and proved
for both the directed setting and the undirected setting of Section
\ref{c}. However, we directly treat only the setting with
one modality of communication; if there are multiple modalities
 of communication then all of the statements and proofs in this
 section apply to each modality separately.
 Some of the results in this section can be found in similar
 or different form in \cite{STFP}; we include all necessary results
 for completeness, and in order to incorporate many substantive
 changes needed for treatment of this article's focus.

\begin{lemma} It almost always holds that $\| AA^T -\X \Y^T(\X \Y^T)^T\|_F \leq \sqrt{3}n^{3/2}\sqrt{\log n}$
and  it almost always holds that $\| A^TA -(\X \Y^T)^T\X \Y^T\|_F \leq \sqrt{3}n^{3/2}\sqrt{\log n}$.
\label{aa}
\end{lemma}
\noindent {\bf Proof:} Let $\X_{i,\cdot}$ and $\Y_{i,\cdot}$ denote the $i$th
rows of $\X$ and $\Y$, respectively. For all $i \ne j$,
\begin{eqnarray} \label{ii}
[AA^T]_{ij}-[\X \Y^T(\X \Y^T)^T]_{ij}=\sum_{l \ne i,j}(A_{il}A_{jl}-\X_{i,\cdot}
\Y_{l,\cdot}^T \X_{j,\cdot} \Y_{l,\cdot}^T)-\X_{i,\cdot}
\Y_{i,\cdot}^T \X_{j,\cdot} \Y_{i,\cdot}^T -\X_{i,\cdot}
\Y_{j,\cdot}^T \X_{j,\cdot} \Y_{j,\cdot}^T
\end{eqnarray}
Hoeffding's inequality states that if $\Upsilon$ is the sum of
$m$ independent random variables that take values in the interval $[0,1]$,
and if $c>0$
then $\Pry \left [ (\Upsilon-\textup{E}[\Upsilon])^2\geq c \right ] \leq 2e^{-\frac{2c}{m}}$.
Thus, for all $i,j$ such that $i \ne j$, if we condition on $\X$ and $\Y$, we have for $l\ne i,j$
that the $m:=n-2$ random variables $A_{il}A_{jl}$ have distribution
Bernoulli$(\X_{i,\cdot} \Y_{l,\cdot}^T \X_{j,\cdot} \Y_{l,\cdot}^T)$
and are independent. Thus,
taking $c=2(n-2)\log n$ in Equation (\ref{ii}), we obtain that
\begin{eqnarray} \label{jj}
\Pry \left [ ([AA^T]_{ij}-
[\X \Y^T( \X \Y^T)^T]_{ij})^2 \geq 2(n-2)\log n + 4n-4 \right ] \leq \frac{2}{n^{4}}.
\end{eqnarray}
Integrating Equation (\ref{jj}) over $\X$ and $\Y$ yields that Equation (\ref{jj})
is true unconditionally. By probability subadditivity, summing over $i,j$ such that $i \neq j$
in Equation (\ref{jj}), we obtain that
\begin{eqnarray} \label{kk}
\Pry \left  [ \sum_{i,j:i \neq j} ([AA^T]_{ij}-[\X \Y^T(\X \Y^T)^T]_{ij})^2
\geq 2n(n-1)(n-2)\log n + 4n(n-1)^2  \right ] \leq \frac{2n(n-1)}{n^4}.
\end{eqnarray}
By the Borel-Cantelli Lemma (which states that if a
sequence of events have probabilities with bounded sum
then almost always the events do not occur) we obtain from
Equation (\ref{kk}) that almost always
\begin{eqnarray*}
\sum_{i,j:i \neq j} ([AA^T]_{ij}-[\X \Y^T(\X \Y^T)^T]_{ij})^2
\leq \frac{5}{2}n^3 \log n
\end{eqnarray*}
and thus almost always $\| AA^T-\X \Y^T(\X \Y^T)^T \|^2_F \leq  3n^3 \log n$ because
each of the diagonals of  $AA^T-\X \Y^T( \X \Y^T)^T$ are bounded in absolute value by $n$. The very same argument holds mutatis mutandis for $\|A^TA-(\X \Y^T)^T \X \Y^T\|^2_F$.~$\qed$\\

The next lemma, Lemma \ref{bb},
provides bounds on the singular values
$\varsigma_1,\varsigma_2,\varsigma_3,\ldots$ of matrix
$\X \Y^T$ and then, in Corollary \ref{cc}, we obtain
bounds on the singular values $\sigma_1,\sigma_2,\sigma_3,\ldots$ of matrix $A$.
Recall that the rank of $\X \Y^T$ is (almost always) $\textup{rank}M$, while
$A$ may in fact have rank $n$.

\begin{lemma}
It almost always holds that $\alpha \gamma n \leq \varsigma_{\textup{rank}M}$, and it always holds
that $\varsigma_1\leq n$. \label{bb}
\end{lemma}
\noindent {\bf Proof:}
Because $\X \Y^T$ is in $[0,1]^{n \times n}$, the nonnegative
matrix $\X \Y^T(\X \Y^T)^T$
has all of its entries bounded by $n$, thus all of its
row sums bounded by $n^2$,
and thus its spectral radius $\varsigma_1^2$ is bounded by $n^2$,
ie we have $\varsigma_1\leq n$ as desired.\\
Next, for all $k=1,2,\ldots,K$, let random variable $n_k$ denote the number
of vertices in block $k$.
The nonzero eigenvalues of $(\X \Y^T)(\X \Y^T)^T=\X \Y^T \Y \X^T$ are
the same as the nonzero eigenvalues of  $\Y^T \Y \X^T \X$.
By the definition of $\alpha$ and the Law of Large Numbers,
almost always $n_k > \alpha n$ for each $k$, thus we express
$\X^T \X=\sum_{k=1}^K n_k \mu_{k,\cdot}^T \mu_{k,\cdot} = \alpha n \mu^T \mu + \sum_{k=1}^K(n_k-\alpha n)\mu_{k,\cdot}^T \mu_{k,\cdot}$
as the sum of two positive semidefinite matrices and obtain that
the minimum eigenvalue of $\X^T\X$ is at least $\alpha \gamma n$.
Similarly the minimum eigenvalue of $\Y^T \Y$ is at least $\alpha \gamma n$.
The minimum eigenvalue of a product of positive semidefinite matrices
is at least the product of their minimum eigenvalues \cite{ZZ},
thus the minimum eigenvalue of $\Y^T \Y \X^T X$ (which is equal to
$\varsigma^2_{\textup{rank}M}$) is at least
 $\alpha \gamma n \cdot \alpha \gamma n$, as~desired.~$\qed$

\begin{corollary} It almost always holds that
$\alpha \gamma n \leq \sigma_{\textup{rank}M}$, it always
holds that $\sigma_1 \leq n$, and
it almost always holds that $\sigma_{\textup{rank}M+1}
\leq 3^{1/4} n^{3/4} \log^{1/4}n$. \label{cc}
\end{corollary}

\noindent {\bf Proof:} By Lemma \ref{aa} and Weyl's Lemma (e.g., see \cite{HJ}), we obtain that for all $m$ it almost always holds that
$| \sigma^2_m-\varsigma^2_m| \leq \| AA^T -\X \Y^T(\X \Y^T)^T\|_F
\leq \sqrt{3}n^{3/2}\sqrt{\log n}$. For all $m > \textup{rank}M$,
the $m$th singular value of $\X \Y^T$ is zero, thus
almost always $\sigma_{\textup{rank}M+1}\leq 3^{1/4} n^{3/4} \log^{1/4}n$.
Lemma \ref{bb} can in fact be strengthened
to show that there is an $\delta >0$ such that almost always
$(\alpha \gamma +\delta) n \leq \varsigma_{\textup{rank}M}$, hence
$(\alpha \gamma +\delta)^2 n^2 \leq \varsigma_{\textup{rank}M}^2$, thus we have
almost always that $(\alpha \gamma)^2 n^2 \leq
\sigma_{\textup{rank}M}^2$, as desired. Showing
that $\sigma_1 \leq n$ is done the same way that $\varsigma_1\leq n$
was  shown in Lemma \ref{bb}. $\qed$\\

It is worth noting that a consequence of Corollary \ref{cc} is that,
for any chosen real number $\omega$ such that $\frac{3}{4}<\omega <1$,
the random variable which counts the number of $\sigma_1,\sigma_2,
\ldots,\sigma_n$ which are greater than $n^\omega$ is a consistent
estimator for $\textup{rank}M$
(is almost always equal to $\textup{rank}M$). Our goal in this
article is to show a robustness result, that ``overestimating"
$\textup{rank}M$ with $R$ in the adjacency-spectral partitioning
procedure does not ruin the consistency of the procedure.\\

Recall from Section \ref{b} the singular value decomposition
$A=[U | U_r ] (\Sigma \oplus \Sigma_r) [V|V_r]^T$. At this point it
will useful to further partition $U=[U_\ell|U_c]$,
$V=[V_\ell|V_c]$, and $\Sigma=\Sigma_\ell \oplus \Sigma_c$ where
$U_\ell,V_\ell \in \R^{n \times \textup{rank}M}$,
$U_c,V_c \in \R^{n \times (R-\textup{rank}M)}$, $\Sigma_\ell \in
\R^{\textup{rank}M \times \textup{rank}M}$, and $\Sigma_c \in
\R^{(R-\textup{rank}M) \times (R-\textup{rank}M)}$.
(The subscripts $\ell,c,r$ are mnemonics for ``left", ``center", and ``right",
respectively.) Also define the matrices $X_\ell := U_\ell \sqrt{\Sigma_\ell}$,
$Y_\ell := V_\ell \sqrt{\Sigma_\ell}$, $X_c := U_c \sqrt{\Sigma_c}$,
$Y_c := V_c \sqrt{\Sigma_c}$, $X_r := U_r \sqrt{\Sigma_r}$, and
$Y_r := V_r \sqrt{\Sigma_r}$. Referring back to the
definition of $X$ and $Y$ in Section \ref{b}, note that
$X=[X_\ell | X_c]$ and $Y=[Y_\ell | Y_c]$.

From the definition of $\beta$ in Section \ref{g} if follows that
for any $i$ and $j$ such that $\X_{i,\cdot} \neq \X_{j,\cdot}$
(or $\Y_{i,\cdot} \neq \Y_{j,\cdot}$) it holds
that $\| \X_{i,\cdot} - \X_{j,\cdot} \| \geq \beta$ (respectively,
$\| \Y_{i,\cdot} - \Y_{j,\cdot} \| \geq \beta$ ). The next result
shows how this separation extends to the rows of the singular vectors of
$\X \Y^T$.

\begin{lemma} Almost always the following are true:\\
For all  $i,j$ such that $\|\X_{i,\cdot}-\X_{j,\cdot}\|_2 \geq \beta$,
it holds that $\| \UU_{i, \cdot} - \UU_{j, \cdot } \|_2 \geq
\beta \sqrt{\frac{\alpha \gamma}{n}}$.\\
For all  $i,j$ such that $\|\Y_{i,\cdot}-\Y_{j,\cdot}\|_2 \geq \beta$,
it holds that $\| \VV_{i, \cdot} - \VV_{j, \cdot } \|_2 \geq
\beta \sqrt{\frac{\alpha \gamma}{n}}$.\\
For all  $i,j$ such that $\|\X_{i,\cdot}-\X_{j,\cdot}\|_2 \geq \beta$,
it holds that
$\| \UU_{i, \cdot}Q \sqrt{\Sigma_\ell} - \UU_{j, \cdot }Q
\sqrt{\Sigma_\ell} \|_2 \geq \alpha \beta \gamma $ for any orthogonal matrix
$Q \in \R^{\textup{rank}M \times \textup{rank}M}$.\\
For all  $i,j$ such that $\|\Y_{i,\cdot}-\Y_{j,\cdot}\|_2 \geq \beta$,
it holds that
$\| \VV_{i, \cdot}Q \sqrt{\Sigma_\ell} - \VV_{j, \cdot }Q
\sqrt{\Sigma_\ell} \|_2 \geq \alpha \beta \gamma$ for any orthogonal matrix
$Q \in \R^{\textup{rank}M \times \textup{rank}M}$.
\label{ff}
\end{lemma}

\noindent {\bf Proof:}
Recall the singular value decomposition
${\mathcal X}{\mathcal Y}^T={\mathcal U} \Lambda {\mathcal V}^T$
from Section \ref{g} (where
${\mathcal U},{\mathcal V} \in \R^{n \times \textup{rank}M}$
each have orthonormal columns and $\Lambda \in \R^{\textup{rank}M
\times \textup{rank}M}$ is diagonal).
Let $\Y^T \Y=W \Delta^2 W^T$ be a spectral decomposition; that is,
$W \in \R^{\textup{rank}M \times \textup{rank}M}$ is orthogonal and
$\Delta \in \R^{\textup{rank}M \times \textup{rank}M}$ is a diagonal matrix
with positive diagonal entries.
Note that
\begin{eqnarray} \label{ll}
(\X W \Delta)(\X W \Delta)^T= \X W \Delta^2 W^T \X^T = \X \Y^T \Y \X^T=\UU
\Lambda \VV^T \VV \Lambda \UU^T=(\UU \Lambda)(\UU \Lambda)^T.
\end{eqnarray}
For any $i,j$ distinct, let $e \in \R^n$ denote the vector with all
zeros except for the value $1$ in the $i$th coordinate and
the value $-1$ in the $j$th coordinate. By Equation (\ref{ll}), we thus have that
\begin{eqnarray} \label{mm}
\|( \X W \Delta)_{i,\cdot}-(\X W \Delta)_{j,\cdot}\|_2^2=
e^T( \X W \Delta)(\X W \Delta)^Te =
e^T (\UU \Lambda )( \UU \Lambda)^T  e = \| ( \UU \Lambda )_{i,\cdot}
-( \UU \Lambda )_{j,\cdot}\|_2^2.
\end{eqnarray}
From Lemma \ref{bb} and its proof, we have that the
diagonals of $\Delta$ are almost always at least
$\sqrt{\alpha \gamma n}$ and that the diagonals of $\Lambda$
are at most $n$. Using this and Equation (\ref{mm}),
 we get that if $i,j$ are such that $\|X_{i,\cdot}-X_{j,\cdot}\| \geq \beta$ then
it holds that
\begin{eqnarray*}
\beta \leq \| \X_{i,\cdot}- \X_{j,\cdot} \|_2=
\| (\X W)_{i,\cdot}-(\X W)_{j,\cdot}\|_2 \leq
\frac{1}{\sqrt{\alpha \gamma n}}
\| (\X W\Delta)_{i,\cdot} - (\X W\Delta)_{j,\cdot} \|_2 \\
= \frac{1}{\sqrt{\alpha \gamma n}}
\|(\UU \Lambda )_{i,\cdot}-(\UU \Lambda)_{j,\cdot} \|_2
\leq  \frac{1}{\sqrt{\alpha \gamma n}} n
\| \UU_{i,\cdot}-\UU_{j,\cdot} \|_2.
\end{eqnarray*}
Thus $\| \UU_{i, \cdot} - \UU_{j, \cdot } \|_2 \geq
\beta \sqrt{\frac{\alpha \gamma}{n}}$, as desired.
Now, if $Q \in \R^{\textup{rank}M \times \textup{rank}M}$ is any orthogonal
matrix then, by Corollary \ref{cc},
\begin{eqnarray*}
\| \UU_{i,\cdot}- \UU_{j,\cdot} \|_2
= \| \UU_{i,\cdot}Q- \UU_{j,\cdot} Q\|_2
\leq \frac{1}{\sqrt{\alpha \gamma n}}
\| \UU_{i, \cdot}Q \sqrt{\Sigma_\ell} - U_{j, \cdot }Q
\sqrt{\Sigma_\ell} \|_2
\end{eqnarray*}
which, together with $\| \UU_{i, \cdot} - \UU_{j, \cdot } \|_2 \geq
\beta \sqrt{\frac{\alpha \gamma}{n}}$, implies
$\| \UU_{i, \cdot}Q \sqrt{\Sigma_\ell} - \UU_{j, \cdot }Q
\sqrt{\Sigma_\ell} \|_2 \geq \alpha \beta \gamma $, as desired.
The same argument applies mutatis mutandis for
$\|Y_{i,\cdot}-Y_{j,\cdot} \| \geq \beta$. $\qed$ \\

In the following, the sum of vector subspaces will refer
to the subspace consisting of all sums of vectors from the
summand subspaces; equivalently, it will be
the smallest subspace containing all of the summand subspaces.
The following theorem is due to Davis and Kahan \cite{DK}
in the form presented in \cite{RCY}.

\begin{theorem} \textup{(Davis and Kahan)}
Let $H,H' \in \R^{n \times n}$ be symmetric,
suppose ${\mathcal S} \subset \R$ is an interval, and suppose for
some positive integer $d$ that
${\mathcal W}, {\mathcal W}' \in {\R^{n \times d}}$
are such that the columns of ${\mathcal W}$ form an
orthonormal basis for the sum of the eigenspaces of
$H$ associated with the eigenvalues of $H$ in ${\mathcal S}$, and
the columns of ${\mathcal W}'$ form an
orthonormal basis for the sum of the eigenspaces of
$H'$ associated with the eigenvalues of $H'$ in ${\mathcal S}$.
Let $\delta$ be the minimum distance between
any eigenvalue of $H$ in ${\mathcal S}$ and any
eigenvalue of $H$ not in ${\mathcal S}$.
Then there exists an orthogonal matrix ${\mathcal Q} \in \R^{d \times d}$ such that
$\|{\mathcal W}{\mathcal Q} -{\mathcal W}' \|_F \leq \frac{\sqrt{2}}{\delta}\|H-H'\|_F$.
 \label{dd}
\end{theorem}

\begin{corollary} There almost always exist
real orthogonal matrices $Q_\UU,Q_\VV \in
\R^{\textup{rank}M \times \textup{rank}M}$
which satisfy $\|\UU Q_\UU - U_\ell \|_F \leq
\frac{\sqrt{6}}{\alpha^2 \gamma^2} \cdot \sqrt{ \frac{\log n }{n}}$
and $\|\VV Q_\VV - V_\ell \|_F \leq
\frac{\sqrt{6}}{\alpha^2 \gamma^2} \cdot \sqrt{ \frac{\log n }{n}}$.
Furthermore, it holds that
$\| \tilde{\X}_\ell - X_\ell \|_F \leq
\frac{\sqrt{6}}{\alpha^2 \gamma^2} \cdot \sqrt{\log n}$
and $\| \tilde{\Y}_\ell - Y_\ell \|_F \leq
\frac{\sqrt{6}}{\alpha^2 \gamma^2} \cdot \sqrt{ \log n }$,
where we define $\tilde{\X}_\ell :=\UU Q_\UU \sqrt{\Sigma_\ell}$
and $\tilde{\Y}_\ell:=\VV Q_\VV \sqrt{\Sigma_\ell}$. \label{ee}
\end{corollary}
\noindent {\bf Proof:}
Take ${\mathcal S}$ in Theorem \ref{dd} to be the interval
$(\frac{1}{2}\alpha^2 \gamma^2 n^2, \infty)$.
By Lemma \ref{bb} and Corollary \ref{cc}, we have almost always that
precisely the greatest $\textup{rank}M$ eigenvalues of each of
$H:=\X \Y^T (\X \Y^T)^T$ and $H':=AA^T$ are in ${\mathcal S}$.
By Lemma \ref{bb}, almost always $\delta \geq \alpha^2 \gamma^2 n^2$
(for the $\delta$ in Theorem \ref{dd}) so, by Lemma \ref{aa},
almost always  $\frac{\sqrt{2}}{\delta}\|H-H'\|_F  \leq
\frac{\sqrt{2}}{\alpha^2 \gamma^2 n^2}\sqrt{3}n^{3/2}\sqrt{\log n}$.
With this, the first statements of Corollary \ref{ee} follow
from the Davis and Kahan Theorem  (Theorem \ref{dd}). The
last statements of  Corollary \ref{ee} follow from postmultiplying
$\UU Q_\UU-U_\ell$ with $\sqrt{\Sigma_\ell}$ and then
using Corollary \ref{cc} and the definition of $X_\ell$.  $\qed$\\

Now, choose $\UU_c \in \R^{n \times (R-\textup{rank}M)}$ and
$\UU_r \in \R^{n \times (n-R)}$ such that
$[\UU | \UU_c | \UU_r] \in \R^{n \times n}$ is an orthogonal
matrix. In particular, note that the columns of $\UU_c$ together with
the columns of $\UU_r$ form an orthonormal basis for the eigenspace
associated with eigenvalue $0$ in the matrix
$H:=\X \Y^T (\X \Y^T)^T$.

\begin{corollary} There almost always exists a real orthogonal matrix
$Q \in \R^{(n-\textup{rank}M) \times (n- \textup{rank}M)}$ such that
$\| \  [\UU_c | \UU_r ] \ Q -  \ [ U_c | U_r ] \ \|_F \leq
\frac{\sqrt{6}}{\alpha^2 \gamma^2} \cdot \sqrt{ \frac{\log n }{n}}$.
Define $\tilde{\X}_c \in \R^{n \times (R-\textup{rank}M)}$ and
$\tilde{\X}_r \in \R^{n \times (n-R)}$ such that
$[\tilde{\X}_c|\tilde{\X}_r]:=[\UU_c|\UU_r]Q
\sqrt{\Sigma_c \oplus \Sigma_r}$. Then
$\| \ [\tilde{\X}_c | \tilde{\X}_r ] \ - \
[ X_c | X_r ] \ \|_F \leq
\frac{3^{1/8}6^{1/2}}{\alpha^2 \gamma^2} \cdot n^{-1/8} \log^{5/8}n$. \label{gg}
\end{corollary}

\noindent {\bf Proof:} The first statement of Corollary \ref{gg} is
proven in the exact manner  that we proved Corollary~\ref{ee},
 except that ${\mathcal S}$ is instead taken to be the
 {\it complement} of $(\frac{1}{2}\alpha^2 \gamma^2 n^2, \infty)$.
The second statement of Corollary \ref{gg} follows by postmultiplying
$[\UU_c | \UU_r ]  Q -  [ U_c | U_r ]$
with $\sqrt{\Sigma_c \oplus \Sigma_r}$ and then using Corollary~\ref{cc} and the
definitions of $X_c$  and $X_r$. $\qed$

\begin{note} Almost always it holds that
$\| \tilde{\X}_c \|_F \leq \sqrt{R-\textup{rank}M} \ 3^{1/8}
n^{3/8}\log^{1/8}n $. \label{hh}
\end{note}

\noindent {\bf Proof:} It is clear (with the matrix $Q$ from
Corollary \ref{gg})
that $[\UU_c| \UU_r ] \ Q$
has orthonormal columns, hence the Froebenius norm of the first
$R-\textup{rank}M$ columns is exactly $\sqrt{R-\textup{rank}M}$.
The result follows from postmultiplying these columns by
$\sqrt{\Sigma_c}$ and using Corollary \ref{cc}. $\qed$

\section{Proof of Theorem \ref{a}, consistency of the adjacency-spectral procedure
of Section \ref{b}  \label{f}}%%%%%%%%%%%%%

In this section we prove  Theorem \ref{a}.
Assuming that the number of blocks $K$ is known and that
an upper bound $R$ is known for $\textup{rank}M$, Theorem \ref{a}
states that, for the adjacency-spectral procedure described in  Section \ref{b},
and for any fixed real number $\epsilon > \frac{3}{4}$,
the number of misassignments
$\min_{\textup{bijections }\pi: \{ 1,2,\ldots,K \}
\rightarrow \{ 1,2,\ldots,K \} } | \{ j=1,2,\ldots,n: \tau(j) \ne
\pi(\hat{\tau}(j))  \} |$ is almost always less than $n^\epsilon$.
We focus first on the scenario where there is a single modality of communication,
and we also suppose for now that it is known that the rows of $M$ are
pairwise nonequal.

First, an observation:
Recall from Section \ref{g} that, for each vertex, the block that
the vertex is a member of via the block {\bf membership}
function $\tau$ is characterized by which of the $K$ distinct-valued rows of $\UU$
the vertex is associated with in $\UU$. In Corollary \ref{ee}, we defined $\tilde{\X}_\ell:=\UU Q_\UU \sqrt{\Sigma_\ell}$. Because $\tilde{\X}_\ell$ is $\UU$ times an
invertible matrix (since $\sqrt{\Sigma_\ell}$ is almost always invertible
by Corollary \ref{cc}), the block that the vertex is truly a member
of  is thus characterized by which of the $K$ distinct-valued rows of
$\tilde{\X}_\ell$ the vertex is associated with in
$\tilde{\X}_\ell$. Also recall that the block which the vertex is
assigned to by the block {\bf assignment} function $\hat{\tau}$ is characterized
by which of the at-most-$K$ distinct-valued rows of ${\mathcal C}$ the vertex is
associated with in ${\mathcal C}$---where
${\mathcal C} \in \R^{n \times R}$ was defined as the
matrix which minimized $\| C-X \|_F$ over all matrices
$C \in \R^{n \times R}$ such that there are at most $K$ distinct-valued
rows~in~$C$.

Denote by $0^{n \times (R - \textup{rank}M)}$ the matrix of zeros
in $\R^{n \times (R - \textup{rank}M)}$.
We next show the following:
\begin{eqnarray} \label{bbb}
\mbox{{\it For any fixed} } \xi > \frac{3}{8}, \mbox{ {\it almost
always it holds that} }
\| {\mathcal C} - [\tilde{\X}_\ell |0^{n \times (R - \textup{rank}M)}] \|_F
\leq n^\xi.
\end{eqnarray}
Indeed, by the definition of
${\mathcal C}$, the fact that
$[\tilde{\X}_\ell |0^{n \times (R - \textup{rank}M)}]$ has $K$ distinct-valued rows,
and the triangle inequality, we have that
\begin{eqnarray} \label{aaa}
\| {\mathcal C} - X \|_F \leq \| \  [\tilde{\X}_\ell
|0^{n \times (R - \textup{rank}M)}]    -X  \|_F
\leq \|  [\tilde{\X}_\ell | \tilde{\X}_c ]   -X  \|_F + \| \tilde{\X}_c \|_F.
\end{eqnarray}
Then, by two uses of the triangle inequality and then Equation (\ref{aaa}), we have
\begin{eqnarray*}
\| {\mathcal C} - [\tilde{\X}_\ell |0^{n \times (R - \textup{rank}M)}] \|_F
& \leq & \| {\mathcal C} - [\tilde{\X}_\ell | \tilde{\X}_c ] \|_F
+ \| \tilde{\X}_c \|_F \\
& \leq &  \| {\mathcal C}-X \|_F + \| X - [\tilde{\X}_\ell | \tilde{\X}_c ] \|_F
+ \| \tilde{\X}_c \|_F \\
& \leq & 2 \cdot  \|  [\tilde{\X}_\ell | \tilde{\X}_c ]   -X  \|_F +
2 \cdot \| \tilde{\X}_c \|_F
\end{eqnarray*}
which, by Corollary \ref{ee}, Corollary \ref{gg}, and Note \ref{hh}, is almost always bounded by
\begin{eqnarray*}
2 \left [ \left (
 \frac{\sqrt{6}}{\alpha^2 \gamma^2} \cdot \sqrt{\log n} \right )^2
 +   \left (   \frac{3^{1/8}6^{1/2}}{\alpha^2 \gamma^2}
 \cdot n^{-1/8} \log^{5/8}n
 \right )^2 \right ] ^{1/2} + 2 R^{1/2}  \ 3^{1/8}
n^{3/8}\log^{1/8}n,
\end{eqnarray*}
which is almost always bounded by
$n^\xi$ for any fixed $\xi > \frac{3}{8}$. Thus Line (\ref{bbb}) is shown.

Now, it easily follows from Line (\ref{bbb}) that
\begin{eqnarray} \mbox{ {\it For any fixed} } \epsilon > \frac{3}{4},
\mbox{ {\it the number of rows of} }
{\mathcal C} - [\tilde{\X}_\ell |0^{n \times (R - \textup{rank}M)}] \nonumber  \\
\mbox{ {\it with Euclidean norm at least} } \frac{\alpha \beta \gamma}{3}
\mbox{ {\it is almost always less than} }
n^\epsilon;  \label{ccc}
\end{eqnarray}
indeed, if this was not true, then
$\|{\mathcal C} - [\tilde{\X}_\ell |0^{n \times (R - \textup{rank}M)}]\|_F
\geq \sqrt{n^\epsilon \left ( \frac{\alpha \beta \gamma}{3} \right )^2}$
would contradict Line \ref{bbb}.

Lastly, form balls $B_1,B_2,\ldots,B_K$ of radius
$\frac{\alpha \beta \gamma}{3}$ about the $K$ distinct-valued rows
of $[\tilde{\X}_\ell |0^{n \times (R - \textup{rank}M)}]$;
by Lemma \ref{ff}, these balls are almost always disjoint.
The number of vertices which the block membership function
$\tau$ assigns to each block is almost always at least $\alpha n$, thus
(by Line (\ref{ccc}) and the Pigeonhole Principle)
almost always each ball $B_1,B_2,\ldots,B_K$ contains exactly
one of the $K$ distinct-valued rows of ${\mathcal C}$. And, for any fixed
$\epsilon > \frac{3}{4}$, the number of
misassignments from $\hat{\tau}$ is thus almost always less
than $n^\epsilon$. Theorem  \ref{a} is now
proven in the scenario where there is a single modality of communication
and it is known that the rows of $M$ are pairwise nonequal.

In the general case where there are multiple modalities of communication
and/or the rows of $M$ are not known to be pairwise nonequal, then
the above proof holds mutatis mutandis (affecting relevant bounds by at most a
constant factor); in place of
$X$  use $Y$ or $[X|Y]$
or $[X^{(1)}|X^{(2)}| \cdots | X^{(S)}]$
or $[Y^{(1)}|Y^{(2)}| \cdots | Y^{(S)}]$
or $[X^{(1)}|X^{(2)}| \cdots | X^{(S)}|Y^{(1)}|Y^{(2)}| \cdots | Y^{(S)} ]$
and in place of $[\tilde{\X}_\ell | \tilde{\X}_c ]$ use
$[\tilde{\Y}_\ell | \tilde{\Y}_c ]$
or $[\tilde{\X}_\ell | \tilde{\X}_c | \tilde{\Y}_\ell | \tilde{\Y}_c ]$
or $[\tilde{\X}^{(1)}_\ell | \tilde{\X}^{(1)}_c | \tilde{\X}^{(2)}_\ell |
\tilde{\X}^{(2)}_c | \cdots | \tilde{\X}^{(S)}_\ell | \tilde{\X}^{(S)}_c ]$,
or $[\tilde{\Y}^{(1)}_\ell | \tilde{\Y}^{(1)}_c | \tilde{\Y}^{(2)}_\ell |
\tilde{\Y}^{(2)}_c | \cdots | \tilde{\Y}^{(S)}_\ell | \tilde{\Y}^{(S)}_c ]$,
or $[\tilde{\X}^{(1)}_\ell | \tilde{\X}^{(1)}_c | \tilde{\X}^{(2)}_\ell |
\tilde{\X}^{(2)}_c | \cdots | \tilde{\X}^{(S)}_\ell | \tilde{\X}^{(S)}_c
| \tilde{\Y}^{(1)}_\ell | \tilde{\Y}^{(1)}_c | \tilde{\Y}^{(2)}_\ell |
\tilde{\Y}^{(2)}_c | \cdots | \tilde{\Y}^{(S)}_\ell | \tilde{\Y}^{(S)}_c ]$,
as appropriate, and similar kinds of adjustments.

\section{Consistent estimation for the number of blocks $K$ \label{e}}%%%%%%%%%%%%%%%%%%%%%%

In this section we provide a consistent estimator $\hat{K}$ for the number of
blocks $K$, if indeed $K$ is not known.
(The only assumption used is
our basic underlying assumption that an
upper bound $R$ is known for $\textup{rank}M$.)

To simplify the notation, in this
section we assume that there is only one modality of communication and we
also assume that it is known that the rows of $M$ are distinct-valued. These
simplifying assumptions do not affect the results we obtain, and the
analysis can be easily generalized to the general case in the same manner
as was done at the end of Section~\ref{f}.

In the adjacency-spectral partitioning procedure from Section
\ref{b}, recall that one of the steps was to compute
${\mathcal C} \in \R^{n \times R}$ which minimized $\| C-X \|_F$ over
all matrices $C \in \R^{n \times R}$ such that there are at most $K$ distinct-valued
rows in $C$. Then the block assignment function $\hat{\tau}$ was defined
as partitioning the vertices into $K$ blocks according to equal-valued
corresponding rows~in~${\mathcal C}$.~Let~us~now generalize the procedure
of Section \ref{b}. Suppose that, for any fixed positive integer
$K'$, we instead compute ${\mathcal C} \in \R^{n \times R}$
which minimizes $\| C-X \|_F$ over
all matrices $C \in \R^{n \times R}$ such that there are at most $K'$
distinct-valued rows in $C$. Then the block assignment function  $\hat{\tau}$
is defined as partitioning the vertices into $K'$ parts (some
possibly empty) according to equal-valued corresponding rows in
${\mathcal C}$. We shall call this adjusted procedure ``the adjacency-spectral partitioning procedure from Section \ref{b} with $K'$ parts."

\begin{theorem} \label{fff}
Let real number $\xi$ such that $\frac{3}{8}<\xi < \frac{1}{2}$ be chosen and fixed.
For the adjacency-spectral procedure from Section \ref{b} with
$K'$ parts, if $K'=K$ then almost always $\| {\mathcal C}-X\|_F \leq n^\xi$,
and if $K'<K$ then almost always
$\| {\mathcal C}-X\|_F > n^\xi$.
\end{theorem}

\noindent {\bf Proof:}
Using Equation (\ref{aaa}), Corollary \ref{ee}, Corollary \ref{gg}, and Note \ref{hh}
in the manner used to prove Line (\ref{bbb}),
we obtain that almost always
$\| \  [\tilde{\X}_\ell |0^{n \times (R - \textup{rank}M)}]-X  \|_F \leq
n^\xi$, and that if $K'=K$ then almost always $\| {\mathcal C}-X\|_F \leq n^\xi$.\\
However, if $K'<K$ then, as we did in Section \ref{f}, consider balls
$B_1,B_2,\ldots,B_K$ of radius $\frac{\alpha \beta \gamma}{3}$
about the $K$ distinct-valued rows
of $[\tilde{\X}_\ell |0^{n \times (R - \textup{rank}M)}]$.
By Lemma~\ref{ff}, these balls are almost always disjoint and, in fact,
their centers are almost always at least $\alpha \beta \gamma$ distance one from the other. By the pigeonhole principle, there is at least one ball
that contains none of the $K'$ distinct-valued rows of ${\mathcal C}$.
Together with the fact that each block almost always  has more
than $\alpha n$ vertices, we obtain almost always that
$\| {\mathcal C} - [\tilde{\X}_\ell |0^{n \times (R - \textup{rank}M)}] \|_F
\geq \sqrt{\alpha n \left ( \frac{\alpha \beta \gamma}{3} \right )^2} $.
Thus, almost always $\| {\mathcal C} - X \| \geq
\| {\mathcal C} - [\tilde{\X}_\ell |0^{n \times (R - \textup{rank}M)}] \|_F
- \| \  [\tilde{\X}_\ell |0^{n \times (R - \textup{rank}M)}]-X  \|_F>n^\xi$. $\qed$\\

Let real number $\xi$ such that $\frac{3}{8}<\xi < \frac{1}{2}$ be chosen and fixed.
Define the random variable $\hat{K}$ to be the least positive integer
$K'$ such that for the adjacency-spectral procedure from Section \ref{b} with
$K'$ parts it happens that $\| {\mathcal C}-X\|_F \leq n^\xi$. By Theorem \ref{fff},
we have the following consistency  result for $\hat{K}$.

\begin{theorem} \label{ggg}
Almost always $\hat{K}=K$.
\end{theorem}

\section{The extended adjacency-spectral partitioning procedure \label{hhh}}%%%%%%%%%%%%%

The adjacency-spectral partitioning procedure of Section \ref{b}
assumed that an integer $R$ was known such that $R \geq \textup{rank}M$,
but it also assumed that the number of blocks $K$ was known. We next
extend the adjacency-spectral partitioning procedure of Section \ref{b}
(we call it ``the extended adjacency-spectral partitioning
procedure") so that it only has the
assumption that an integer $R$ is known such that $R \geq \textup{rank}M$,
and it is not assumed that $K$ is known. The procedure is as follows:

Let real number $\xi$ such that $\frac{3}{8}<\xi < \frac{1}{2}$ be chosen and fixed.
Successively for $K'=1,2,3\ldots$,~do~the spectral partitioning procedure of Section \ref{b} with $K'$
parts until it happens that $\| {\mathcal C} - X \|_F \leq n^\xi$, then
return the $\hat{\tau}$ from the last successive iteration (i.e.~the iteration
where $K'=\hat{K}$).

\begin{theorem} \label{iii}
With the extended adjacency-spectral partitioning procedure,
for any fixed $\epsilon>\frac{3}{4}$,
the number of misassignments $\min_{\textup{bijections }\pi: \{ 1,2,\ldots,K \}
\rightarrow \{ 1,2,\ldots,K \} } | \{ j=1,2,\ldots,n: \tau(j) \ne
\pi(\hat{\tau}(j))  \} |$ is almost always less than $n^\epsilon$.
\end{theorem}

\noindent {\bf Proof:} Indeed, almost always the last value of $K'$ (which is
$\hat{K}$) is equal to $K$ by Theorem~\ref{ggg}, and then
almost always the number of misassignments is less than $n^\epsilon$
by Theorem \ref{a}. $\qed$

\section{Another consistent estimator for $K$ \label{nnn}}%%%%%%%%%%%

In Section \ref{e} we provided the consistent estimator $\hat{K}$
for the number of blocks $K$. It was based on Theorem \ref{fff}, which
contrasted---for the adjacency-spectral procedure
from Section~\ref{b} with $K'$ parts---what would happen
when $K'=K$ versus when $K'<K$.
In this section we are interested in contrasting---for the adjacency-spectral
procedure from Section~\ref{b} with $K'$ parts---what would
happen when $K'=K$ versus when $K'>K$.
This yields another consistent estimator for $K$.

For the adjacency-spectral procedure from Section \ref{b} with $K'$ parts,
the at-most $K'$ distinct-valued rows of ${\mathcal C}$ will be called
the {\it centroids}, the {\it centroid separation} will refer to the minimum
Euclidean distance between all pairs of distinct centroids, and the
{\it minimum part size} will refer to the least cardinality of the
$K'$ parts as partitioned by $\hat{\tau}$; in particular, if one of the parts
is empty then the minimum part size is zero, whereas the centroid
separation would still be positive.

\begin{theorem} \label{mmm}
For the adjacency-spectral procedure from Section \ref{b} with
$K'$ parts, if $K'=K$ then almost always the
minimum part size is greater than $\alpha n$ and the centroid
separation is at least $\frac{\alpha \beta \gamma}{3}$. Let $\zeta>0$ and
$\vartheta >0$ be any fixed real numbers. If $K'>K$ then almost always
it will {\bf not} hold that the minimum part size is
greater than $\vartheta n$ and the centroid separation
is at least $\zeta$. \label{ddd}
\end{theorem}

\noindent {\bf Proof:}
As we did in Section \ref{f}, consider balls
$B_1,B_2,\ldots,B_K$ of radius $\frac{\alpha \beta \gamma}{3}$
about the $K$ distinct-valued rows
of $[\tilde{\X}_\ell |0^{n \times (R - \textup{rank}M)}]$.
By Lemma \ref{ff}, these balls are almost always disjoint and, in fact,
their centers are almost always at least $\alpha \beta \gamma$ distance one from the other.
If  $K=K'$ then recall from Section \ref{f} that almost always
each ball contains exactly one centroid. By the $\alpha \beta \gamma$ separation
between the balls' centers, we thus have almost always that
the centroid separation is at least $\frac{\alpha \beta \gamma}{3}$.
Also, by Theorem \ref{a} there is almost always
 a strictly sublinear number of misassignments, hence almost always
the minimum part size is
greater than $\alpha n$.

Now to the case of $K' > K$. Suppose by way of contradiction that the
minimum part size is greater than $\vartheta n$ and the
centroid separation is at least $\zeta$. Since there are
strictly more centroids than balls $B_1,B_2,\ldots,B_K$, and because of the $\zeta$
separation between the centroids, by the pigeonhole principle
there is at least one centroid with distance greater than $\frac{\zeta}{3}$
from each row of
 $[\tilde{\X}_\ell |0^{n \times (R - \textup{rank}M)}]$ (these rows are
 the centers  of the balls). Since this centroid appears as a
 row of ${\mathcal C}$ more than
 $\vartheta n$ times, this would imply that
 $\|{\mathcal C} - [\tilde{\X}_\ell |0^{n \times (R - \textup{rank}M)}]\|_F
\geq \sqrt{\vartheta n \left ( \frac{\zeta}{3} \right )^2}$. However
we have by the triangle inequality, the definition of ${\mathcal C}$, and the
first few line of the proof of Theorem \ref{fff} that almost always
$\|{\mathcal C} - [\tilde{\X}_\ell |0^{n \times (R - \textup{rank}M)}]\|_F
\leq  \|{\mathcal C} - X \|_F + \| X-
[\tilde{\X}_\ell |0^{n \times (R - \textup{rank}M)}]\|_F
\leq  \|{\mathcal C}_K - X \|_F + \| X-
[\tilde{\X}_\ell |0^{n \times (R - \textup{rank}M)}]\|_F \leq  2n^\xi
< \sqrt{\vartheta n \left ( \frac{\zeta}{3} \right )^2}$
(where $\xi$ such that $\frac{3}{8}<\xi < \frac{1}{2}$ is fixed and
${\mathcal C}_K$ denotes what ${\mathcal C}$ would have been
if we instead did the adjacency-spectral procedure from
Section~\ref{b} with $K$ parts instead of $K'$ parts),
which gives us the desired contradiction.$\qed$ \\

With Theorem \ref{mmm} we obtain another consistent estimator for $K$.
However, we would need to assume that positive real numbers $\zeta$ and
$\vartheta$ are known that satisfy $\vartheta \leq \alpha$ and
$\zeta \leq \frac{\alpha \beta \gamma}{3}$. Assuming that such
$\zeta$ and $\vartheta$ are indeed known, we can define the random variable
$\check{K}$ to be the greatest positive integer $K'$ among the
values $1,2,3,\ldots \lfloor \frac{1}{\vartheta} \rfloor$
(note that $\frac{1}{\vartheta}$ is an upper bound on $K$)
such that for the  adjacency-spectral procedure from Section \ref{b} with
$K'$ parts the minimum part size is
greater than $\vartheta n$ and the centroid separation
is at least $\zeta$.
By Theorem \ref{mmm} we immediately obtain the
following consistency result for $\check{K}$.

\begin{theorem}
Almost always $\check{K}=K$.
\end{theorem}

In order to define $\check{K}$, lower bounds on
$\frac{\alpha \beta \gamma}{3}$ and $\alpha$ need to be known
in addition to an upper bound on $\textup{rank}M$ that needs to be known.
This contrasts
with $\hat{K}$, for which we only need to assume that
an upper bound on $\textup{rank}M$ is known. (Because $\hat{K}$ requires fewer assumptions,
the extended adjacency-spectral partitioning procedure in Section~\ref{hhh}
utilizes $\hat{K}$ and not $\check{K}$.) Nonetheless, it is useful to
be aware of how the adjacency-spectral procedure from Section~\ref{b} with
$K'$ parts changes in behavior when $K'$ becomes greater than $K$---besides
how it changes in behavior when $K'$ becomes less than $K$. And when lower bounds
on $\frac{\alpha \beta \gamma}{3}$ and $\alpha$ are also known then, in practice
for a single value of $n$,
we can check for $\hat{K}=\check{K}$ in order to have more confidence
that their common value is indeed $K$.

\section{A simulated example and discussion \label{ppp}}%%%%%%%%%%%%%%%%%

\begin{figure*}[t]
 \centering
 \includegraphics[width=1.0\textwidth]{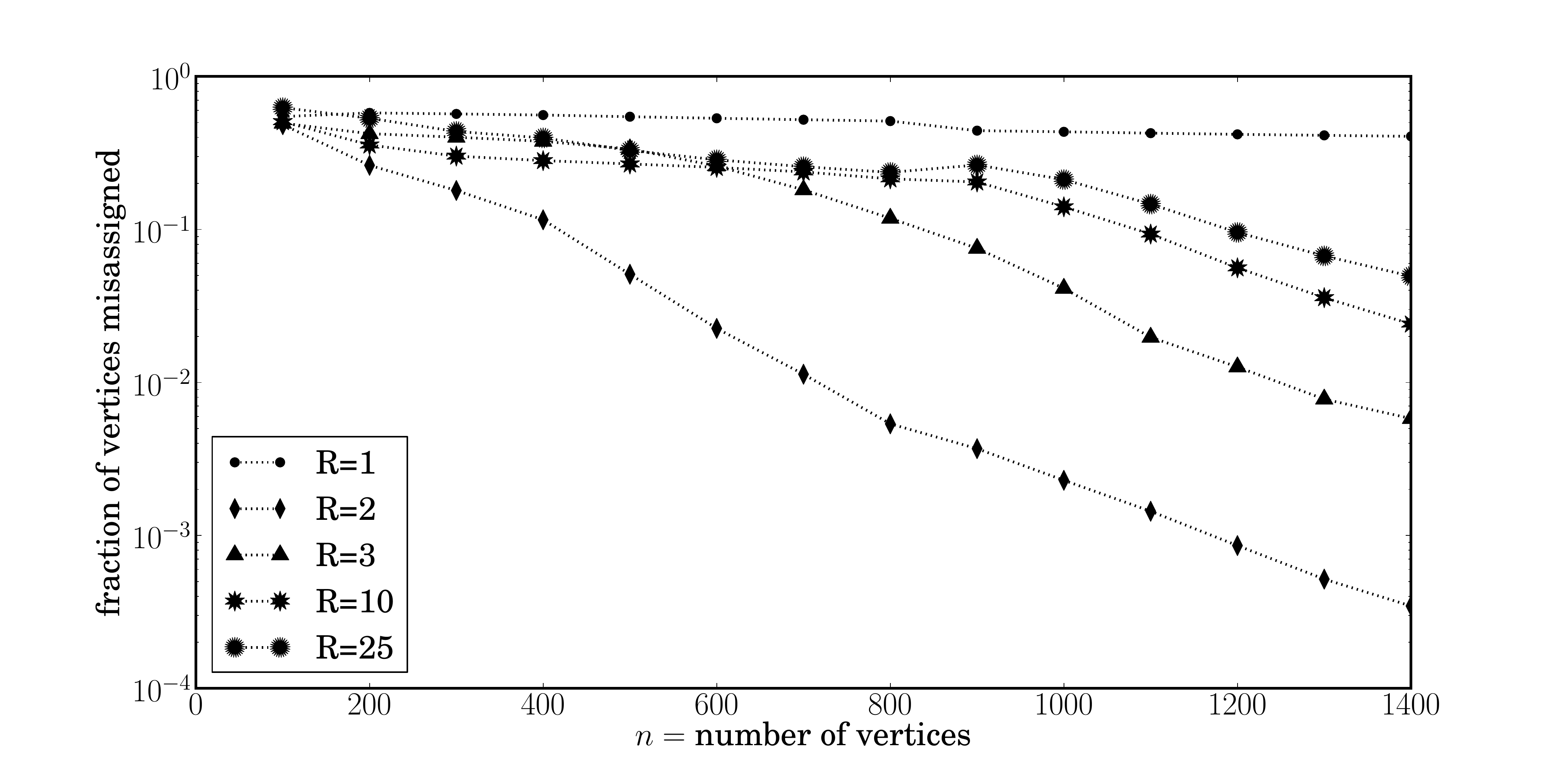}
 \caption{The mean misassignment fraction plotted against $n$,
 for each of $R=1,2,3,10,25$. \label{qqq}}
 \end{figure*}

As an illustration, consider the stochastic block model with parameters
\begin{equation} \label{eq:param1}
K=3 ,\quad
\rho= \left [ \begin{array}{c} .3 \\ .3 \\ .4  \end{array}
\right ] \quad
M=\left [ \begin{array}{ccc}
 .205 & .045 & .150 \\
 .045 & .205 & .150 \\
 .150 & .150 & .180
\end{array}
\right ],
\end{equation}
(in particular, there is only one modality of communication) and suppose
edges are undirected. Here rank$M=2$,

For each of the values $R=1,2,3,10,25$ and for each
number of vertices $n=100,200,300,\ldots,1400$, we generated $2500$
Monte Carlo replications of this stochastic block model and
to each of these $2500$ realizations we applied
the adjacency-spectral partitioning procedure of Section~\ref{b}
using $R$ as the upper bound on rank$M$ (which,
in the case of $R=1$, is purposely incorrect for illustration
purposes) assuming that we know
$K=3$. Note that rather than finding the actual minimum of $\|\mathcal{C}-X\|_F$, we use the $K$-means algorithm which approximates this minimum. The five curves in Figure \ref{qqq} correspond to $R=1,2,3,10,25$
respectively, and they plot the mean fraction of misassignments
(the number of misassigned vertices divided by the total number of vertices~$n$,
such fractions averaged
over the $2500$ Monte Carlo replicates) along the $y$-axis, against
the value of $n$ along the $x$-axis.

Note that when $R=2$ the performance of the
adjacency-spectral partitioning  is excellent
(in fact, the number of misassignments
becomes effectively zero as $n$ gets to $1600$).
Indeed, even when $R=10$ and $R=25$ (which is substantially greater than  rank$M=2$)
the adjacency-spectral partitioning partitioning
performs very well.
However, when $R=1$, which is not an upper bound on rank$M$ (violating
our one assumption in this article), the misassignment rate
of adjacency-spectral partitioning is almost as bad as chance.

\begin{figure}
\begin{center}
\includegraphics[width=\textwidth]{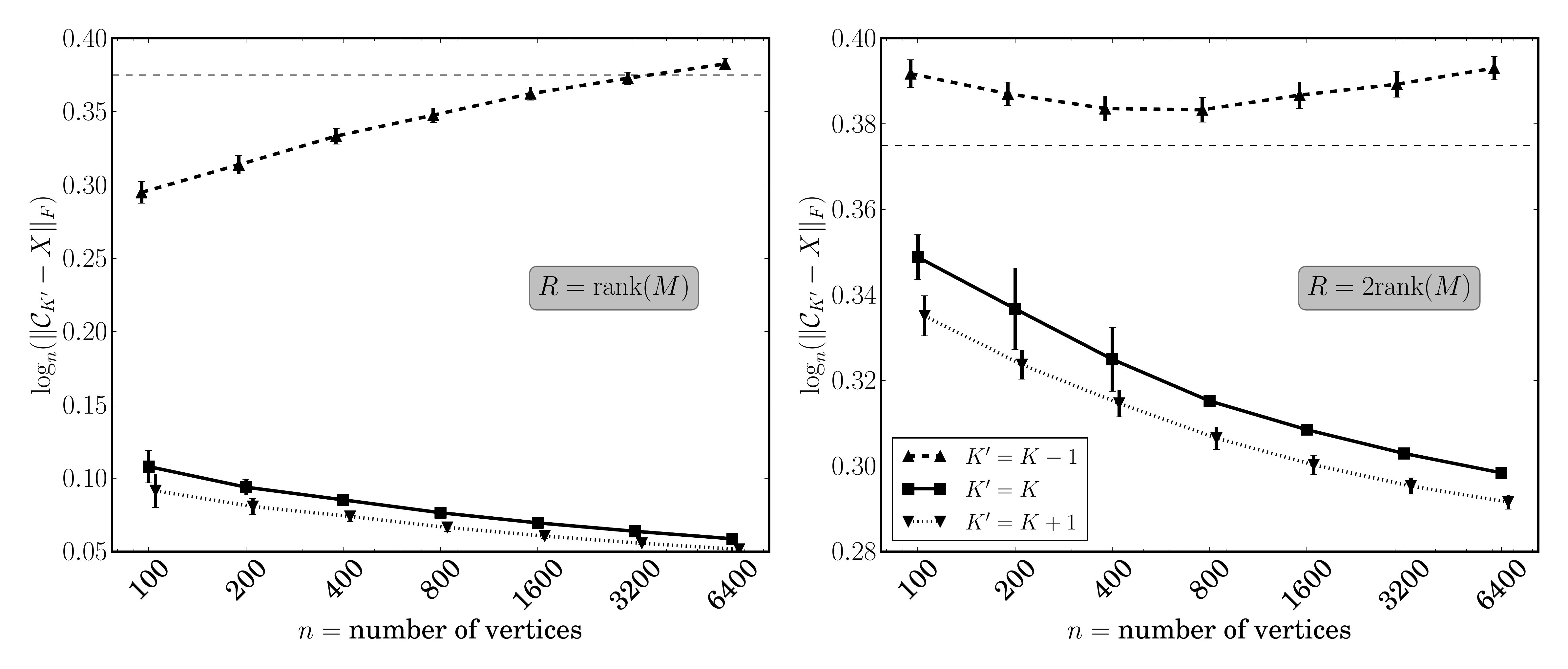} % data at ~\Dropbox\Data\k_est_v5.pickle, plot affiliationSims
\end{center}
\caption{Test statistic for estimating $K$ using the parameters in line~\eqref{eq:kestParam} for $R=3,6$ and $K'=2,3,4$. The unmarked dash line shows $\xi=3/8$.}
\label{fig:kest}
\end{figure}

Next we will consider the estimator for $K$ proposed in Section~\ref{e}. Recall that this estimator is defined as $\hat{K}=\arg\min_{K'} \{\| {\mathcal C}_{K'} - X \|_F \leq n^\xi\}=\arg\min_{K'} \{\log_n (\|\mathcal{C}_{K'}-X\|_F)\leq \xi\}$  where $\mathcal{C}_{K'}$ is the $n\times R$ matrix of centroids associated with each vertex, the adjacency spectral clustering procedure in Section~\ref{b} is done with $K'$ parts, and $\xi\in(3/8,1/2)$ is fixed. We now consider stochastic block model parameters with stronger differences between blocks to illustrate the effectiveness of the estimator. In particular we let
\begin{equation}\label{eq:kestParam}
K=3, \quad \rho= \begin{bmatrix} .3 \\ .3 \\ .4  \end{bmatrix}
 \quad
M=\begin{bmatrix}
 .5 & .1 & .1 \\
 .1 & .5 & .1 \\
 .1 & .1 & .5
\end{bmatrix}
\end{equation}
so that $\mathrm{rank}M=3$. 
For each $n=100,200,400,800,1600,3200,6400$ we generated $50$
Monte Carlo replications of this stochastic block model. 
To each of these $50$ realizations we performed the adjacency spectral clustering procedure using $R=3$ (Figure~\ref{fig:kest}, left panel) and $R=6$ (Figure~\ref{fig:kest}, right panel) as our upper bound but this time assuming $K$ is not known. We used $K'=2,3,4$ and computed the statistic $\log_n (\|\mathcal{C}_{K'}-X\|_F)$. Figure~\ref{fig:kest} shows the mean and standard deviation of this test statistic over the $50$ Monte Carlo replicates for each $R$, $K'$ and $n$.

The results demonstrate that for $n=6400$, $\hat{K}$ is a good estimate when $R=3=\mathrm{rank}M$ when we choose $\xi$ close to $3/8$. On the other hand for smaller values of $n$, our estimator will select too few blocks regardless of the choice of $\xi\in (3/8,1/2)$. 
Interestingly, choosing $\xi$ close to $3/8$, $\hat{K}$ always equals the true number of blocks when we let $R=6=2\mathrm{rank}M$, suggesting that this estimator has interesting behavior as a function of $R$. 
Note that for larger values of $\xi$, $\hat{K}$ will tend to be smaller, and for smaller values of $\xi$, $\hat{K}$ will tend to be larger.

\section{Discussion}\label{sec:disc}
Our simulation experiment for estimating $K$ demonstrates that good performance is possible for moderate $n$ under certain parameter selections. This buttresses the theoretical and practical interest, as this estimator may serve as a stepping stone for the development of other more effective estimators. Indeed, bounds shown in \cite{Oliveira} suggest that it may be possible to allow $\xi$ to be as small as $1/4$ using different proof methods. These methods in terms of the operator norm are an important area for further investigation when considering spectral techniques for inference on random graphs.

Note additionally that for our first simulation, we used $k$-means rather than minimizing $\|\mathcal{C}-X\|_F$ since the latter is computationally unfeasible. This, together with fast methods to compute the singular value decomposition, indicates that this method can be used even on quite large graphs. For even larger graphs, there are also techniques to approximate the singular value decomposition that should be considered in future work.

Further extensions of this work can be made in various directions. Rohe et al. \cite{RCY} and others allow for the number of blocks to grow. We believe that this method could be extended to this scenario, though careful analysis is necessary to show that the estimator for the number of blocks is still consistent. 

Another avenue is the problem of missing data, in the form of missing edges; results for this setting follow immediately provided that the edges are missing uniformly at random. This is because the observed graph will still be a stochastic block model with the same block structure. Other forms of missing data are deserving of further study. Sparse graphs are also of interest and this work can likely be extended to the case of moderately sparse graphs, for example with minimum degree $\Theta(n/\sqrt{\log n})$, without significant additional machinery. 
Another form of missing data is that since we consider graphs with no self-loops, the diagonal of the adjacency matrix are all zeros. Marchette et al. \cite{Marchette} and Scheinerman and Tucker \cite{Scheinerman} both suggest methods to impute the diagonals, and this has been show to improve inference in practice.

This is related to one final point to mention:
Is it better to do spectral partitioning on the adjacency matrix (as we
do here in this article) or on the Laplacian (to be used in place of
the adjacency matrix in our procedure of this article)? There doesn't
currently seem to be a clear answer; for some choices of stochastic block model
 parameters it seems empirically that the
adjacency matrix gives fewer misassignments than
the Laplacian, and for other choices of parameters the Laplacian seems
to be better.
%(Indeed, for the specific example in this section, the
%adjacency matrix seems empirically to be better.)
A determination of exact criterion (on the stochastic block model parameters)
for which the adjacency matrix is better than the Laplacian and vice versa
deserves attention in future work. But the analysis that we
used here to reduce the required knowledge of the model parameters
and to show robustness in the procedure will hopefully serve as an
impetus to achieve formal results for spectral partitioning in the
nonparametric setting for which the block model assumptions don't hold.\\

\noindent {\bf Acknowledgements:} 
This work (all authors) is partially supported by National Security Science and Engineering Faculty Fellowship (NSSEFF) grant number N00244-069-1-0031,
Air Force Office of Scientific Research (AFOSR), and Johns Hopkins University Human Language Technology Center of Excellence (JHU HLT COE). We also thank the editors and the anonymous referees for their valuables comments and critiques that greatly improved this work.

\end{document}